\documentclass[prl,twocolumn,superscriptaddress,showpacs]{revtex4}
\usepackage{graphicx}
\usepackage{dcolumn}
\usepackage{amsmath}
\usepackage{color}

\begin{document}

\title{Structural relaxation due to electronic correlations in the paramagnetic insulator KCuF$_3$}
\author{I. Leonov}
\affiliation{Abdus Salam International Center for Theoretical Physics, Trieste 34014,
Italy}
\author{N. Binggeli}
\affiliation{Abdus Salam International Center for Theoretical Physics, Trieste 34014,
Italy}
\affiliation{INFM-CNR Democritos, Theory @ Elettra group, Trieste 34014, Italy}
\author{Dm. Korotin}
\affiliation{Institute of Metal Physics, S. Kovalevskoy St. 18, 620219 Yekaterinburg
GSP-170, Russia}
\author{V. I. Anisimov}
\affiliation{Institute of Metal Physics, S. Kovalevskoy St. 18, 620219 Yekaterinburg
GSP-170, Russia}
\author{N. Stoji\' c}
\affiliation{International School for Advanced Studies, SISSA, Via Beirut 2/4, 34014
Trieste, Italy}
\affiliation{INFM-CNR Democritos, Theory @ Elettra group, Trieste 34014, Italy}
\author{D. Vollhardt}
\affiliation{Theoretical Physics III, Center for Electronic
Correlations and Magnetism, Institute of Physics, University of
Augsburg, Augsburg 86135, Germany}
\date{\today}

\begin{abstract}
A computational scheme for the investigation of complex materials
with strongly interacting electrons is formulated which is able to
treat atomic displacements, and hence structural relaxation, caused
by electronic correlations. It combines \textit{ab initio} band
structure and dynamical mean-field theory and is implemented in
terms of plane-wave pseudopotentials. The equilibrium Jahn-Teller
distortion and antiferro-orbital order found for paramagnetic
KCuF$_3$ agree well with experiment.
\end{abstract}

 \maketitle

In materials with correlated electrons, the interaction between spin,
charge, orbital, and lattice degrees of freedom leads to a wealth of
ordering phenomena and complex phases \cite{Rev}. The diverse
properties of such systems and their great sensitivity with respect
to changes of external parameters such as temperature, pressure,
magnetic field, or doping also make them highly attractive for
technological applications \cite{Rev}. In particular, orbital
degeneracy is an important and often inevitable cause for this
complexity \cite{KK}. A fascinating example is the cooperative
Jahn-Teller (JT) effect --- the spontaneous lifting of the
degeneracy of an orbital state --- leading to an occupation of
particular orbitals (``orbital ordering'') and, simultaneously, to a
structural relaxation with symmetry reduction.

The electronic structure of materials can often be described quite
accurately by density functional theory in the local density
approximation (LDA) \cite{LDA} or the generalized gradient
approximation (GGA) \cite{PB96,GGALDA}. However, these methods
usually fail to predict the correct electronic and structural
properties of materials where electronic correlations play a role.
Extensions of LDA, e.g., LDA+U \cite{LA95} and SIC-LDA \cite{ZPO80+}
can improve the results, e.g., the band gap value and local moment,
but only for systems with long-range order. Hence the computation of
electronic, magnetic, and structural properties of strongly
correlated \textit{paramagnetic} materials remains a great
challenge. Here the recently developed combination of band structure
approaches and dynamical mean-field theory \cite{DMFT}, the
so-called LDA+DMFT computational scheme \cite{DMFTmeth}, has become
a powerful new tool for the investigation of strongly correlated
compounds in both their paramagnetic and magnetically ordered
states. This technique has recently provided important insights into
the properties of correlated electron materials \cite{DMFTcalc},
especially in the vicinity of a Mott metal-insulator transition as
encountered in transition metal oxides \cite{Rev}.

Applications of  LDA+DMFT so far mainly
employed linearized and higher-order muffin-tin orbital [L(N)MTO]
methods \cite{LNMTO} and concentrated on the study of correlation
effects within the electronic system for a given ionic lattice. On
the other hand, the interaction of the electrons with the ions also
affects the lattice structure. LDA+DMFT investigations of
particularly drastic examples, the volume collapse in paramagnetic
Ce \cite{HM01,AB06} and Pu \cite{SKA01+} and the magnetic moment
collapse in MnO \cite{MnO}, incorporated the lattice by calculating
the total energy of the correlated material as a function of the
atomic volume. However, for investigations going beyond equilibrium
volume calculations, e.g., of the cooperative JT effect and other
subtle structural relaxation effects, the L(N)MTO method is not
suitable since it cannot determine atomic displacements reliably.
This is partly
due to the fact that the atomic-sphere approximation used in the
L(N)MTO scheme, with a spherical potential inside the atomic sphere,
completely neglects multipole contributions to the electrostatic
energy originating from the distorted charge density distribution
around the atoms. By contrast, the plane-wave pseudopotential
approach employed here does not neglect such contributions and can
thus fully describe the effect of the distortion on the
electrostatic energy.

In this Letter, we present a computational scheme which allows us to
calculate lattice relaxation effects caused by electronic
correlations. To this end, the GGA+DMFT --- a merger of the GGA and
DMFT --- is formulated within a plane-wave pseudopotential approach
\cite{TL08,PSEUDO,LG06}. Thereby the limitations of the L(N)MTO
scheme in the direct calculation of total energies are overcome. In
particular, we apply this new method to determine the orbital order
and the cooperative JT distortion in the paramagnetic phase of the
prototypical JT system KCuF$_{3}$.


KCuF$_{3}$ has long been known to be a prototypical material with a
cooperative JT distortion \cite{KK} where the electronic degrees of
freedom are the driving force behind the orbital order
\cite{KK,LA95,MK02}. Indeed, the relatively high (tetragonal)
symmetry makes KCuF$_{3}$ one of the simplest systems to study. In
particular, only a single internal structure parameter, the shift of
the in-plane fluorine atom from the Cu-Cu bond center, is needed to
describe the lattice distortion. Moreover, there is only a single
hole in the $d$-shell so that complications due to multiplet effects
do not arise.
KCuF$_3$\ is an insulating pseudocubic perovskite whose structure
is related to that of high-$T_c$ superconductors and colossal
magnetoresistence manganites.
 The copper ions have octahedral fluorine surrounding and
are nominally in a Cu$^{2+}$ ($3d^9$) electronic configuration, with
completely filled $t_{2g}$ orbitals and a single hole in the $e_g$
states. The cubic degeneracy of the Cu $e_g$ states is lifted due to
a cooperative JT distortion leading to an elongation of the CuF$_6$
octahedra along the $a$ and $b$ axis, and an antiferro--distortive
pattern in the $ab$ plane \cite{BM90}. This is associated with an
alternating occupation of $d_{x^2-z^2}$ and $d_{y^2-z^2}$ hole
orbitals along the $a$ and $b$ axes, resulting in a tetragonal
compression ($c/a < 1$) of the unit cell. Purely electronic effects
as in the Kugel-Khomskii theory \cite{KK} and the electron-lattice
\cite{G63} interaction have been discussed as a possible mechanism
behind the orbital ordering in KCuF$_3$. The antiferro ($a$-type)
and ferrolike ($d$-type) stacking of the $ab$ planes along the $c$
axis give rise to two different structural polytypes, which have
been identified experimentally at room temperature \cite{O69}. 

Below the N\' eel temperature ($T_N$$\sim$38 K for $a$-type and $\sim$22 K for $%
d$-type ordering), which is much lower than the critical temperature
for orbital ordering, KCuF$_3$ shows $A$-type antiferromagnetic
 order \cite{HS69}.
The antiferromagnetic structure is consistent with the
Goodenough-Kanamori-Anderson rules for a superexchange interaction
with $d_{x^2-z^2}$/$d_{y^2-z^2}$ antiferro-orbital ordering. This is
also found within LDA+U which finds the correct orbitally ordered,
antiferromagnetic insulating ground state \cite{LA95,BA04}, while
LDA predicts \textit{metallic} behavior. Moreover, LDA+U
calculations for a model structure of KCuF$_3$\ in which cooperative
JT distortions are completely neglected reproduce the correct
orbital order, suggesting an electronic origin of the ordering
\cite{LA95,MK02} in agreement with the Kugel-Khomskii theory
\cite{KK}. Altogether, LDA+U is able to determine the JT distortion
in KCuF$_3$ rather well \cite{LA95,BA04}, but simultaneously
predicts an additional long-range magnetic order. Therefore LDA+U
cannot explain the properties at temperatures above $T_N$ and, in
particular, at room temperature, where KCuF$_3$  is a correlated
paramagnetic insulator with a robust JT distortion which persists up
to the melting temperature. To determine the correct orbital order
and cooperative JT distortion for a correlated paramagnet, i.e., to
perform a structural optimization, we here employ GGA+DMFT.

We first calculate the GGA band structure of KCuF$_3$ at room
temperature (space group $I4/mcm$) \cite{BM90}, employing the
plane-wave pseudopotential approach \cite{PSEUDO,calc_details}.
Calculations are performed for values of the in-plane JT distortion
${\delta_{JT}}$ \cite{difJT} ranging from 0.2\% to 7\% while keeping
the  lattice parameters $a $ and $c$ and the space group symmetry
fixed. In the paramagnetic phase, and for all values of
$\delta_{JT}$ considered here, the GGA yields a
 \textit{metallic} rather than the experimentally observed insulating behavior, with an
appreciable orbital polarization due to the crystal field splitting.
Overall, the GGA results qualitatively agree with previous
band-structure calculations \cite{LA95,BA04}. Obviously, a JT distortion
by itself, without the inclusion of electronic correlations in the
paramagnetic phase, cannot explain the experimentally observed
orbitally ordered \emph{insulating} state of KCuF$_3$.

To include the electronic correlations, we construct an effective
low-energy Hamiltonian ${\hat H_{GGA}}$ for the partially filled Cu
$e_g$ orbitals for each value of the distortion $\delta_{JT}$
considered here. This is achieved by employing the pseudopotential
plane-wave GGA results and making a projection onto atomic-centered
symmetry-constrained Cu $e_g$ Wannier orbitals \cite{TL08}. Taking
the local Coulomb repulsion $U $ and Hund's rule exchange $J$ into
account, one obtains the following low-energy Hamiltonian for the
two ($m=1,2$) Cu $e_g$ bands:
\begin{eqnarray}
{\hat H} & = & {\hat H_{GGA}} + U\sum_{im} n_{im\uparrow} n_{im\downarrow}
\notag \\
& + & \sum_{i \sigma \sigma^{\prime}} (V - \delta_{\sigma
\sigma^{\prime}} J) n_{i1\sigma} n_{i2\sigma^{\prime}} - {\hat
H_{DC}}.
\end{eqnarray}
Here the second and third terms on the right-hand side describe the
local Coulomb interaction between Cu $e_g$ electrons in the same and
in different orbitals, respectively, with $V=U-2J$, and ${\hat
H_{DC}}$ is a double counting correction which accounts for the
electronic interactions already described by the GGA (see below).
To compute the electronic correlation induced structural relaxation
of KCuF$_3$, we calculate the total energy as \cite{AB06,LG06}
\begin{equation}
E = E_{GGA}[\rho] + \langle H_{GGA} \rangle - \sum_{m,k}
\epsilon^{GGA}_{m,k}  + \langle H_{U} \rangle - E_{DC},
\end{equation}
where $E_{GGA}[\rho]$ is the total energy obtained by GGA. The third
term on the right-hand side of Eq.~(2) is the sum of the GGA Cu
$e_g$ valence-state eigenvalues and is given by the thermal average
of the GGA Hamiltonian with the GGA Green function $G^{GGA}_{\bf
k}(i\omega_n)$:
\begin{equation}
\sum_{m,k} \epsilon^{GGA}_{m,k} = \frac{1}{\beta}~\sum_{n,{\bf k}}
Tr[H_{GGA}(\mathbf{k}) G^{GGA}_{\mathbf{k}}(i\omega_n)]
e^{i\omega_n0^{+}}.
\end{equation}

$\langle H_{GGA} \rangle$ is evaluated similarly but with the full
Green function including the self-energy. The interaction energy
$\langle H_{U} \rangle$ is computed from the double occupancy
matrix. The double-counting correction $E_{DC}= \frac{1}{2}U
N_{e_g}(N_{e_g}- 1)-\frac{1}{4}J N_{e_g}(N_{e_g}- 2)$
corresponds to the average Coulomb repulsion between the $N_{e_g}$
electrons in the Cu $e_g$ Wannier orbitals.

The many-body Hamiltonian (1) is solved within DMFT for $U=7$ eV and
$J=0.9$ eV \cite{LA95}  using quantum Monte Carlo (QMC) calculations
\cite{HF86,temperature,Off-diag-elements}.
\begin{figure}[tbp!]
\centerline{\includegraphics[width=0.35\textwidth,clip]{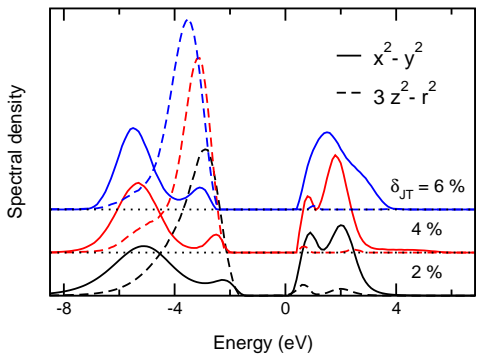}}
\caption{(colour online) Orbitally resolved Cu $e_g$ spectral
densities of paramagnetic KCuF$_3$ as obtained by GGA+DMFT(QMC) for
different values of the JT distortion. } \label{fig:spectra}
\end{figure}
Figure~\ref{fig:spectra} shows the spectral density of paramagnetic
KCuF$_3$, obtained from the QMC data by the maximum entropy method,
for three values of the JT distortion $\delta_{JT}$. Most
importantly, a paramagnetic insulating state with a strong orbital
polarization is obtained for all $\delta_{JT}$.
 The energy gap is in the
range 1.5--3.5 eV, and increases with increasing $\delta_{JT}$. The
sharp feature in the spectral density at about $-3$ eV corresponds
to the fully occupied $3z^2-r^2$ orbital \cite{dif_orb}, whereas the
lower and upper Hubbard bands are predominantly of $x^2-y^2$
character and are located at $-5.5$ eV and 1.8 eV, respectively.
\begin{figure}[tbp!]
\centerline{\includegraphics[width=0.35\textwidth,clip]{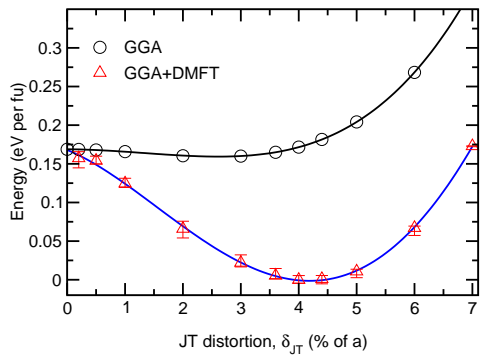}}
\caption{(colour online) Comparison of the total energies of
paramagnetic KCuF$_3$ computed by GGA and GGA+DMFT(QMC) as a
function of the JT distortion. Error bars indicate the statistical
error of the DMFT(QMC) calculations. } \label{fig:energy}
\end{figure}

The total energies as a function of the JT distortion obtained by
the GGA and GGA+DMFT, respectively, are compared in
Fig.~\ref{fig:energy}. We note that the GGA not only predicts a
\textit{metallic} solution, but its total energy is seen to be
almost constant for $0<\delta_{JT}\lesssim 4\%$. Both features are
in contradiction to experiment since the extremely shallow minimum
at $\delta_{JT}\simeq2.5\%$ would imply that KCuF$_3$ has no JT
distortion for $T\gtrsim$ 100 K. By contrast, the inclusion of the
electronic correlations among the partially filled Cu $e_g$ states
in the GGA+DMFT approach leads to a very substantial lowering of the
total energy by $\sim$ 175 meV per formula unit. This implies
that the strong JT distortion persists up to the melting temperature
($>1000$ K), in agreement with experiment. The minimum of the
GGA+DMFT total energy is located at the value $\delta_{JT}= 4.2\%$,
which is also in excellent agreement with the experimental value of
4.4\% \cite{BM90}. This clearly shows that the JT distortion in
paramagnetic KCuF$_3$ is caused by electronic correlations.

\begin{figure}[tbp!]
\centerline{\includegraphics[width=0.4\textwidth,clip]{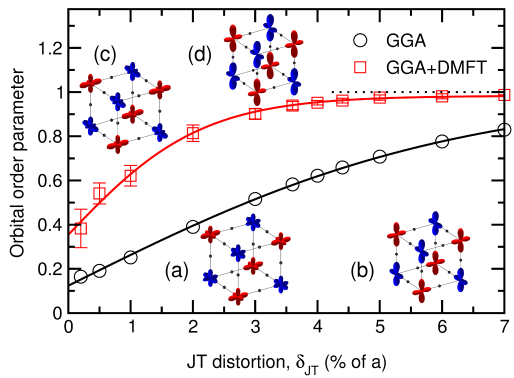}}
\caption{(colour online) Dependence of the orbital order parameter
in paramagnetic KCuF$_3$ on the JT distortion as obtained by GGA and
GGA+DMFT(QMC), respectively. Error bars indicate the statistical
error of the DMFT(QMC) calculations. Insets (a)/(b) refer to GGA and
(c)/(d) to GGA+DMFT results and show the hole orbital ordering for
$\delta_{JT}=$0.2\%/4.4\% (see text). } \label{fig:order_par}
\end{figure}

An analysis of the occupation matrices for the $e_g$ Cu Wannier
states obtained by the  GGA+DMFT calculations confirms a substantial
orbital
polarization in the calculated paramagnetic phase of KCuF$_3$. As shown in Fig.~\ref%
{fig:order_par}, the orbital order parameter (defined as the
difference between $3z^2-r^2$ and $x^2-y^2$ Cu $e_g$ Wannier
occupancies \cite{dif_orb}) saturates at about 98\% for 
$\delta_{JT} \gtrsim$ 4\%. Thus, the GGA+DMFT result shows 
a predominant occupation of the Cu $3z^2-r^2$ orbitals.
We note that even without a JT distortion the orbital order
parameter would remain quite large ($\sim$40\%). Moreover, while the
GGA result for $\delta_{JT}= 0$  yields a symmetric orbital
polarization with respect to $C_4$ rotations around the $c$ axis,
spontaneous antiferro-orbital order is found in GGA+DMFT. This
difference is illustrated in Fig.~\ref{fig:order_par}, where insets
(a) and (c) depict the hole orbital order obtained  by the GGA and
GGA+DMFT for $\delta_{JT}=0.2$\%, respectively. The GGA charge
density is more or less the same along the $a$ and $b$ axis [inset
(a)]; i.e.,  the Cu $d_{x^2-z^2}$ and $d_{y^2-z^2}$ hole orbitals
are almost equally occupied and hence are not ordered. By contrast,
the GGA+DMFT results clearly show an alternating occupation [inset
(c)], corresponding to the occupation of a $x^2-y^2$ hole orbital in
the local coordinate system, which implies antiferro-orbital order.
For the experimentally observed value of the JT distortions of
$\delta_{JT}=4.4$\%, both GGA and GGA+DMFT find antiferro-orbital
order [insets (b),(d)]. However, we note again that, in contrast to
the GGA+DMFT, the GGA yields a \textit{metallic} solution without
any JT distortion for $T\gtrsim$ 100 K, in contradiction to
experiment.

In conclusion, by formulating GGA+DMFT ---  the combination of the
\emph{ab initio} band structure calculation technique GGA with the
dynamical mean-field theory --- in terms of plane-wave
pseudopotentials \cite{TL08}, we constructed a robust computational
scheme for the investigation of complex materials with strong
electronic interactions. Most importantly, this framework is able to
determine the correlation induced structural relaxation of a solid.
Results obtained for paramagnetic KCuF$_3$, namely an equilibrium
Jahn-Teller distortion of 4.2\% and antiferro-orbital ordering,
agree well with experiment. The electronic correlations were also
found to be responsible for a considerable enhancement of the
orbital polarization.
The GGA+DMFT scheme presented in this paper opens the way for fully
microscopic investigations of the structural properties of strongly
correlated electron materials such as lattice instabilities observed
at correlation-induced metal-insulator transitions.

\begin{acknowledgments}
We thank M. Altarelli, J. Deisenhofer, D. Khomskii, S. Streltsov, and G.
Trimarchi for valuable discussions. Support by the
Russian Foundation for Basic Research under Grant No.
RFFI-07-02-00041, the Deutsche Forschungsgemeinschaft through SFB
484, and the Light Source Theory Network, LighTnet of the EU is
gratefully acknowledged.
\end{acknowledgments}


\begin{thebibliography}{99}
\bibitem{Rev} M. Imada, A. Fujimori, and Y. Tokura, Rev. Mod.
Phys. \textbf{70}, 1039 (1998); Y. Tokura and N. Nagaosa, Science \textbf{288%
}, 462 (2000); E. Dagotto, Science \textbf{309}, 257 (2005).

\bibitem{KK}
K. I. Kugel and D. I. Khomskii, Sov. Phys. Usp. \textbf{25} (4), 231
(1982).


\bibitem{LDA}  R. O. Jones and O. Gunnarsson, Rev. Mod. Phys. {\bf 61},
689 (1989).

\bibitem{PB96} J. P. Perdew, K. Burke, and M. Ernzerhof, Phys. Rev. Lett.
\textbf{77}, 3865 (1996).

\bibitem{GGALDA} In general, GGA tends to give better results than
LDA for the electronic and structural properties of complex oxides
and related materials. See, D. R. Hamann, Phys. Rev. Lett.
\textbf{76}, 660 (1996) and H. Sawada \emph{et al.},
Phys. Rev. B \textbf{56}, 12154 (1997).


\bibitem{LA95} A. I. Liechtenstein, A. I. Anisimov, and J. Zaanen, Phys.
Rev. B \textbf{52}, 5467 (1995).

\bibitem{ZPO80+} J. P. Perdew and A. Zunger, Phys. Rev. B
\textbf{23}, 5048 (1981).

\bibitem{DMFT} W. Metzner and D. Vollhardt, Phys. Rev. Lett. \textbf{62},
324 (1989); A. Georges \emph{et al.},
Rev. Mod. Phys. \textbf{68}, 13 (1996); G. Kotliar and D. Vollhardt,
Phys. Today \textbf{57}, No. 3, 53 (2004); G. Kotliar \emph{et al.},
Rev. Mod. Phys. \textbf{78}, 865 (2006).

\bibitem{DMFTmeth} V. I. Anisimov \emph{et al.}, J. Phys. Condens. Matt. \textbf{9}%
, 7359 (1997); A. I. Lichtenstein and M. I. Katsnelson, Phys. Rev. B \textbf{%
57}, 6884 (1998); K. Held \textit{et al.}, Psi-k Newsletter \textbf{56}, 65
(2003); K. Held \textit{et al.}, Phys. Status Solidi B \textbf{243}, 2599
(2006).

\bibitem{DMFTcalc} K. Held \emph{et al.},
Phys. Rev. Lett. \textbf{86}, 5345 (2001); E. Pavarini \emph{et
al.},
Phys. Rev. Lett. \textbf{92}, 176403 (2004); A. I. Poteryaev, A. I.
Lichtenstein, and G. Kotliar, Phys. Rev. Lett. \textbf{93}, 086401
(2004); S. Biermann \emph{et al.},
Phys. Rev. Lett. \textbf{94}, 026404 (2005); L. Chioncel \emph{et
al.}, Phys. Rev. Lett. \textbf{96}, 197203 (2006); J. Kunes
\emph{et al.},
Phys. Rev. Lett. \textbf{99}, 156404 (2007).

\bibitem{LNMTO} O. K. Andersen, Phys. Rev. B \textbf{12}, 3060 (1975);
O. K. Andersen and T. Saha-Dasgupta, Phys. Rev. B \textbf{62}, R16219 (2000).


\bibitem{HM01}
A. K. McMahan, K. Held, and R. T. Scalettar, Phys. Rev. B
\textbf{67}, 075108 (2003).

\bibitem{AB06} B. Amadon \emph{et al.},
Phys. Rev. Lett. \textbf{96}, 066402 (2006).

\bibitem{SKA01+} S. Y. Savrasov, G. Kotliar, and E. Abrahams, Nature
(London) \textbf{410}, 793 (2001); X. Dai \emph{et al.},
Science \textbf{300}, 953 (2003); S. Y. Savrasov and G. Kotliar,
Phys. Rev. B \textbf{69}, 245101 (2004).

\bibitem{MnO} J. Kunes \emph{et al.},
Nature Materials 7, 198 (2008).

\bibitem{TL08} G. Trimarchi \emph{et al.}, J. Phys.: Condens. Matter
\textbf{20}, 135227 (2008).
Dm. Korotin \emph{et al.}, cond-mat/08013500.

\bibitem{PSEUDO} Calculations have been done using the PWSCF package:
S. Baroni \emph{et al.},
URL http://www.pwscf.org.

\bibitem{LG06} For a formulation of LDA+DMFT within a
mixed-basis pseudopotential approach see F. Lechermann \emph{et
al.}, Phys. Rev. B \textbf{74}, 125120 (2006).


\bibitem{MK02} J. E. Medvedeva \emph{et al.},
Phys. Rev. B \textbf{65}, 172413 (2002).

\bibitem{BM90} R. H. Buttner, E. N. Maslen, and N. Spadaccini, Acta Cryst. B
\textbf{46}, 131 (1990).

\bibitem{G63} J. B. Goodenough, \textit{Magnetism and the Chemical Bond}
(Interscience, New York, 1963).

\bibitem{O69} A. Okazaki, J. Phys. Soc. Jpn. \textbf{26}, 870 (1969);
\textbf{27}, 518B (1969).

\bibitem{HS69} M. T. Hutchings \emph{et al.},
Phys. Rev. \textbf{188}, 919 (1969).

\bibitem{BA04} N. Binggeli and M. Altarelli, Phys. Rev. B \textbf{70},
085117 (2004).


\bibitem{calc_details}
Calculations were performed with the Perdew-Burke-Ernzerhof
exchange-correlation functional \cite{PB96} together with Vanderbilt
ultrasoft pseudopotentials for Cu and F, a soft Troullier-Martin
pseudopotential for K, and a kinetic energy cutoff of 75 Ry for the
plane-wave expansion of the electronic states.

\bibitem{difJT} We define the Jahn-Teller distortion by 
$\delta_{JT}= \frac{1}{2}(d_l-d_s)/(d_l+d_s)$.
Here $d_l$ and $d_s$ denote the long and short Cu-F bond distances,
respectively. The  structural data \cite{BM90} at room-temperature
yield $\delta_{\mathrm{JT}}= 4.4$\% (in units of the lattice
constant $a$).

\bibitem{HF86} J. E. Hirsch and R. M. Fye, Phys. Rev. Lett \textbf{56}, 2521
(1986).

\bibitem{temperature} Calculations were performed at $T=1160$ K to make
the QMC simulations \cite{HF86} feasible.
In the present study this is not an important
limitation since there is no structural transitions above 300 K.

\bibitem{Off-diag-elements} To simplify the computation we neglected
the orbital off-diagonal elements of the local Green function by
applying an additional transformation into the local basis set with
a diagonal density matrix during each DMFT iteration.

\bibitem{dif_orb} The local coordinate system is chosen with the \emph{z}-direction
defined along the longest Cu-F bond of the CuF$_6$ octahedron.


\end{thebibliography}
\end{document}